\documentclass[prl,twocolumn,floatfix,altaffilletter,superscriptaddress,preprintnumbers,tightenlines, showkeys]{revtex4-1}

\pdfoutput=1

\usepackage[utf8]{inputenc}
\usepackage[colorlinks=true,citecolor=blue,linkcolor=blue]{hyperref}
\usepackage[normalem]{ulem}
\usepackage{amsmath,amssymb}
\usepackage{epsfig}
\usepackage{graphicx}              
\usepackage{url} 
\usepackage{color}
\usepackage{slashed}
\usepackage{multirow}
\usepackage{placeins}
\usepackage[dvipsnames]{xcolor}
\usepackage{epstopdf}
\usepackage{soul} 
\usepackage{tikz}
\usepackage{mathtools}
\usepackage{xcolor}
\usepackage{amsbsy}

\def\beq{\begin{equation}}
\def\eeq{\end{equation}}
\def\bea{\begin{eqnarray}}
\def\eea{\end{eqnarray}}
\def\ben{\begin{enumerate}}
\def\een{\end{enumerate}}

\def\lsim{\mathrel{\raise.3ex\hbox{$<$\kern-.75em\lower1ex\hbox{$\sim$}}}}
\def\gsim{\mathrel{\raise.3ex\hbox{$>$\kern-.75em\lower1ex\hbox{$\sim$}}}}
\def\ifmath#1{\relax\ifmmode #1\else $#1$\fi}

\def\GeV{~{\mbox{GeV}}}

\def\simlt{\stackrel{<}{{}_\sim}}

\def\nn{\nonumber}

\def\mI{\mathcal{I}}

\def\mL{\mathcal{L}}

\def\1{\textbf{1}}
\def\2{\textbf{2}}
\def\3{\textbf{3}}
\def\4{\textbf{4}}
\def\5{\textbf{5}}
\def\6{\textbf{6}}
\def\7{\textbf{7}}
\def\8{\textbf{8}}
\def\9{\textbf{9}}

\newcommand{\Eq}[1]{Eq.~(\ref{#1})}

\begin{document}
%====================================================================%

\title{A  $\pmb{\nu}$ Solution to the Strong CP Problem}

\author{Marcela Carena}
\email{carena@fnal.gov}
\affiliation{Fermi National Accelerator Laboratory, P.~O.~Box 500, Batavia, IL 60510, USA}
\affiliation{Enrico Fermi Institute and Kavli Institute for Cosmological Physics, University of Chicago, Chicago, IL 60637, USA}
\author{Da Liu}
\email{da.liu@anl.gov}
\affiliation{HEP Division, Argonne National Laboratory, 9700 Cass Ave., Argonne, IL 60439, USA}
\author{Jia Liu}
\email{liuj1@uchicago.edu}
\affiliation{Enrico Fermi Institute and Kavli Institute for Cosmological Physics, University of Chicago, Chicago, IL 60637, USA}
\author{Nausheen~R.~Shah}
\email{nausheen.shah@wayne.edu}
\affiliation{Department of Physics $\&$ Astronomy, Wayne State University, Detroit, MI 48201, USA}
\author{Carlos~E.~M.~Wagner}
\email{cwagner@anl.gov}
\affiliation{Enrico Fermi Institute and Kavli Institute for Cosmological Physics, University of Chicago, Chicago, IL 60637, USA}
\affiliation{HEP Division, Argonne National Laboratory, 9700 Cass Ave., Argonne, IL 60439, USA}
\author{Xiao-Ping Wang}
\email{xia.wang@anl.gov}
\affiliation{HEP Division, Argonne National Laboratory, 9700 Cass Ave., Argonne, IL 60439, USA}

\preprint{
FERMILAB-PUB-19-144-T
\\\phantom{0} \hfill EFI-19-4
\\\phantom{0} \hfill WSU-HEP-1902}

%====================================================================%

\begin{abstract}
We present a new solution to the strong CP problem in which the imaginary component of the up quark mass, $\mathcal{I}[m_u]$, acquires a tiny, but non-vanishing value.   This is achieved via a Dirac seesaw mechanism, which is also responsible for the generation of the small neutrino masses. Consistency with the observed value of the up quark mass is achieved via instanton contributions arising from QCD-like interactions. In this framework, the value of the neutron electric dipole moment  is directly related to $\mathcal{I}[m_u]$, which, due to its common origin with the neutrino masses, implies that the neutron electric dipole moment is likely to be measured in the next round of experiments. We also present a supersymmetric extension of this Dirac seesaw model to stabilize the hierarchy among the scalar mass scales involved in this new mechanism.
\end{abstract}

\maketitle
\flushbottom

%====================================================================%

\noindent\underline{\textbf{\textit{Introduction.}}}  The Standard Model (SM) has been highly successful in describing all experimental observations~\cite{Tanabashi:2018oca}. The observed flavor and CP-violating effects originate from the weak interactions via the dependence of the charged currents on the unitary CKM matrix $V_{\rm CKM}$. There is, however, another potential source of CP violation in the SM, associated with the strong interaction. After the diagonalization of the quark masses, the QCD Lagrangian density contains the terms
\bea \label{eq:QCD_lag}
{\cal L} & \supset &  - \frac{\theta \  g_s^2}{32 \pi^2} G_{\mu\nu,a} \tilde{G}^{\mu\nu,a}-  \sum_q  \left(m_q \bar{q}_L q_R + h.c.\right),
\eea
where $g_s$ is the strong gauge coupling, $G_{\mu \nu, a}$ is the QCD field strength tensor,  $\tilde{G}_{\mu\nu,a} =\frac{1}{2}\epsilon_{\mu\nu\alpha\beta} G^{\alpha\beta,a}$ is its dual, and $m_q$ are the quark masses. Due to the QCD chiral anomaly, the value of $\theta$ can be modified by a phase redefinition of the chiral quark fields, but the physical value 
\begin{equation}
\theta_{\rm QCD} = \theta + {\rm arg}\left[ \det[ M_q] \right],
\label{eq:thetaQCD}
\end{equation}
where $\det [M_q] = \prod m_q$, remains invariant. As will be discussed in detail later on, a non-vanishing value of $\theta_{\rm QCD}$ leads to QCD induced CP-violating effects, like the neutron electric dipole moment (nEDM), which is as yet unobserved. The current bound on the nEDM, $d_n < 3.0 \times10^{-26} {\rm e \, cm}$~\cite{Afach:2015sja,Baker:2006ts}, leads to the constraint $\theta_{\rm QCD}(1\GeV)  \simlt 1.3 \times 10^{-10}$. The dynamical origin of such small values of $\theta_{\rm QCD}$ is the so-called strong CP problem. 

The $\theta$ term in Eq.~(\ref{eq:QCD_lag}) may be eliminated by a proper phase redefinition of the quark fields. For a non-zero $\theta_{\rm QCD}$, at least one of the quark masses, for instance the up quark mass, would become a complex quantity, with argument $\theta_{\rm QCD}\sim \mI[m_u]/|m_u|$. 
Hence in such a case, all the QCD-induced CP-violating effects would  be associated with  $\mathcal{I}[m_u]$, and  would vanish in the limit of zero  up quark mass. This is the well known massless up quark solution to the strong CP problem~\cite{Georgi:1981be, Kaplan:1986ru,Choi:1988sy, Banks:1994yg, Davoudiasl:2007zx, Dine:2014dga,  Bardeen:2018fej}.

We shall denote as the {\it canonical basis}, the basis in which $\theta = 0$ and $\theta_{\rm QCD}$ is the argument of the up quark mass. Using the value of the up quark mass determined in the framework of chiral perturbation theory, 
$|m_u({\rm 1~GeV})| \simeq 5~{\rm MeV}$~\cite{Gasser:1982ap}, the bound on $\theta_{\rm QCD}$ becomes equivalent to
\begin{equation}
\mathcal{I}[m_u (1~{\rm GeV})] \simlt 6.5 \times 10^{-4} {\rm eV}\; .
\label{eq:bound1}
\end{equation} 
The relevant question then becomes, can one dynamically generate a value of 
$\mathcal{I}[m_u(1~{\rm GeV})]$ consistent with such a stringent bound, while the real part, $\mathcal{R}[m_u(1~{\rm GeV})]$, is of the order of a few MeV? 

To analyze this question, one should remember that the up quark mass at scales of the order of 1~GeV receives contributions not only from  its tree-level Higgs Yukawa interaction, which we will denote as $m_u^H$, but  also from  instanton contributions, $m_u^{\rm inst}$. Hence in general,
\begin{equation}
m_u(1~{\rm GeV}) =  m_u^{\rm inst} + m_u^H.
\end{equation}
In the case of QCD, the instanton contributions to the up quark mass depend on the masses of the other quarks in the theory.  In a general basis, the light quark 
contributions are given by~\cite{Georgi:1981be,Choi:1988sy},
\begin{equation}
 m_u^{\rm inst}  = \frac{\exp(- i \theta) \left(m_d^H m_s^H \right)^*}{\Lambda},
 \label{eq:muinst}
\end{equation}
where $\Lambda$ is a scale which characterizes the size of these contributions, and $m_d^H$ and $m_s^H$ are the tree-level Higgs induced down and strange quark masses. 

In the canonical basis, $m_u^{\rm inst}$  is a real contribution,  implying that $\mathcal{I}[m_u^H] = \mathcal{I}[m_u]$.  The physical CP-violating phase, Eq.~(\ref{eq:thetaQCD}), then reads
\beq
\label{eq:thetaQCD1GeV}
\theta_{\rm QCD} (1 \GeV) \simeq \sin \theta^{\rm H}_{\rm QCD} \ \frac{|m_u^H|}{|m_u|}(1 \GeV),
\eeq
where $\theta^{\rm H}_{\rm QCD} = {\rm arg}[m_u^H]$ and we have assumed that $|m_u^H| \ll |m_u^{\rm inst}|$.  This expression is consistent with $\theta_{\rm QCD} = {\rm arg}[m_u]$. The small imaginary components of the instanton induced strange and down quarks masses, proportional to $m_d^H (m_u^H)^*/\Lambda$ and $m_s^H (m_u^H)^*/\Lambda$ respectively, induce a subdominant effect that becomes negligible in the one instanton approximation.     
From  Eqs.~(\ref{eq:bound1}) and (\ref{eq:thetaQCD1GeV}), we conclude that  a strong CP problem solution would be provided if values of 
\begin{equation}
|m_u^H (1~{\rm GeV})|   \sin \theta^{\rm H}_{\rm QCD}   \simlt 6.5 \times 10^{-4} {\rm eV}\;, 
\label{eq:bound2}
\end{equation} 
could be dynamically generated while maintaining consistency with the observed up quark mass.

Interestingly, it has been  argued that the $m_u^{\rm inst}$ contribution induced by the standard QCD interactions may be as large as a few~MeV~\cite{Georgi:1981be, Kaplan:1986ru, Choi:1988sy, Banks:1994yg, Davoudiasl:2007zx, Dine:2014dga,  Bardeen:2018fej}, and hence be able to explain the observed up quark mass value. This is allegedly in tension with the lattice determination of the up quark mass at scales where the instanton contribution should be negligible, namely,  $m_u^{\overline{\rm MS}}(2~{\rm GeV}) = 2.1$~MeV~\cite{Bazavov:2018omf, Aoki:2019cca}. Alternatively, it has been postulated that similar contributions may come from instantons in some ultraviolet gauge extensions~\cite{ Agrawal:2017evu}.  A possible ultraviolet configuration is that each generation is sensitive to a different $SU(3)$ gauge interaction,  with a gauge group $SU(3)^3=SU(3)_1 \times SU(3)_2 \times SU(3)_3$ that is spontaneously broken to the diagonal group $SU(3)$ at a scale of the order of hundreds of TeV.  Assuming that the tree-level Higgs induced strange and bottom quark masses are equal to zero, the instanton contributions in each sector would be responsible for bringing these masses to their observed values (via contributions proportional to the charm and top quark masses, respectively). In such a case, the low energy CP-violating interactions will be governed by  expressions similar to \Eq{eq:thetaQCD1GeV}, with the only difference that $m_u^{\rm inst}$ will include the ultraviolet instanton contributions.  
Hence, any  tension of the up quark mass with lattice determinations would be eliminated.

Irrespective of its origin, provided $m_u^{\rm inst}$ can lead to the observed up quark mass at scales of the order of 1~GeV,  $m_u^H$ can be arbitrarily  small. One would naturally expect the  Higgs induced CP-violating phase  $\theta^{\rm H}_{\rm QCD}$ to be larger than $\sim10^{-2}$. In such a case, from Eq.~(\ref{eq:bound2}), $|m_u^H|$ would  be of the order of or smaller than $4 \times 10^{-2}$~eV.  This implies values of $|m_u^H|$  similar in magnitude to the small  neutrino masses~\cite{Tanabashi:2018oca}.    
Our proposed solution of the strong CP problem is associated with the dynamical generation of precisely such small values of $|m_u^H|$.  \\

%====================================================================%

\begin{figure}
	\centering
	\includegraphics[width=0.5 \columnwidth]{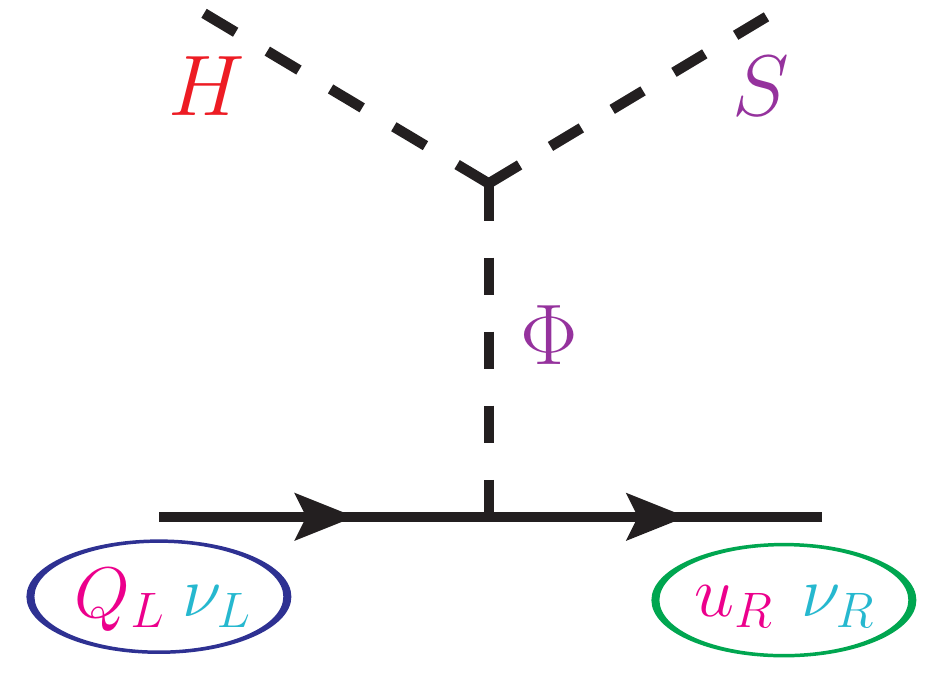} 
	\caption{A diagrammatic representation of the Dirac seesaw mechanism for the up quark and neutrino masses. }
	\label{fig:dirac-seesaw}
\end{figure}

%====================================================================%

\noindent\underline{\textbf{\textit{A Dirac Seesaw Model.}}} We present a model which realizes a seesaw mechanism for the dynamical generation of $m_u^H$ and of small Dirac neutrino masses~\cite{Gu:2006dc,Bonilla:2017ekt}. To realize this idea, we assume the presence of a $\mathcal{Z}_4$ discrete symmetry that forbids the direct coupling of the up quark and neutrinos to the Higgs field. While the right-handed up quark and the right-handed neutrinos have charge 1, all  other SM fields carry zero charge under this symmetry.  In addition, we introduce a heavy scalar doublet $\Phi$ with $\mathcal{Z}_4$ charge 1 and hypercharge $1/2$, and  a singlet $S$ of charge -1 under the $\mathcal{Z}_4$ symmetry, such that  $\nu_R$ Majorana masses  are forbidden.  

The Lagrangian for the Yukawa interactions of the up quark and the neutrinos is given by:
\begin{equation}
\mL = Y_\nu \bar{\ell}_L\tilde{\Phi} \nu_R  + Y_u \bar{q}_L  \tilde{\Phi} u_R  + h.c.,
\end{equation}
where $\tilde{\Phi} = i \sigma_2 \Phi^*$ carries charge -1 under $\mathcal{Z}_4$. All the other SM fermions have  standard Yukawa interactions with the Higgs doublet $H$, which are not shown here. The  potential involving the heavy scalar fields relevant for our discussion reads
\begin{eqnarray}
\label{eq:potential}
&V &=  m^2_{\Phi} \Phi^\dagger \Phi   + (\rho  S H^\dagger \Phi +h.c.)  +  \lambda_{\Phi,1}  \Phi^\dagger \Phi    H^\dagger H \nn \\
&+&  \lambda_{\Phi,2}  \Phi^\dagger \Phi | S |^2 + \lambda_{S,1}  |S|^4 + (\lambda_{S,2} S^4 + h.c) + \cdots  .
\end{eqnarray}
Here the term $\lambda_{S,2} S^4$ is allowed by the discrete symmetry but would not be allowed by a global  Peccei-Quinn $U(1)$ symmetry~\cite{Peccei:1977hh}. Hence, there is no axion-like Goldstone boson~\cite{Wilczek:1977pj,Weinberg:1977ma}.  It is easy to prove that to ensure a vacuum expectation value~(vev) in the real direction and stability of the potential, we need $\lambda_{S,2} <0$ and $(\lambda_{S,1} + \lambda_{S,2}) > 0$.  We will assume that $m_{\Phi} \gg  m_S, m_H$, so that one can integrate it out by the equation of motion $\Phi \simeq - \frac{1}{m_\Phi^2} \rho S^* H$,
where we have assumed that $\rho$ is real.  The effective Yukawa interactions for the up quark and neutrinos, represented  in Fig.~\ref{fig:dirac-seesaw}, are given by:
\beq
\mL_{\rm eff} \simeq - Y_\nu \frac{\rho}{m_\Phi^2} S  \bar{\ell}_L\tilde{H} \nu_R   - Y_u  \frac{\rho}{m_\Phi^2} S  \bar{q}_L \tilde{H} u_R  + h.c. .
\eeq
After the singlet and the neutral component of the  SM Higgs field acquire vevs, $\left<S\right> = v_S/\sqrt{2}, \left<H^0\right> = v/\sqrt{2}$,  the Dirac masses of the up quark and neutrinos read:
\beq
m_\nu \sim Y_\nu \frac{\rho ~v_S v}{2m_\Phi^2}, \qquad m_u^H \sim Y_u \frac{\rho~ v_S v}{2m_\Phi^2} .
\label{eq:YuYnu}
\eeq
If $v_S$ is the order  of the EW scale $v = 246$ GeV, and $\rho$ is of order $m_\Phi$, one gets an effective seesaw suppression of the  up quark and neutrino masses $|m_u|, |m_\nu| \sim v^2/m_\Phi$. Hence, as assumed, one 
sees the need for large values of~$m_\Phi$,
\begin{equation}
m_\Phi \simeq  6\times 10^{12}~{\rm GeV}  \left( \frac{Y_\nu}{0.1} \right) \left( \frac{\rho}{0.1 m_\Phi} \right) \left( \frac{v_S}{v} \right) \left(\frac{0.05~ \rm eV}{m_\nu} \right),
\label{eq:mphi}
\end{equation}
to get an observational consistent mass for the heavier neutrino, where we have assumed $Y_\nu$ to be real.  Given the bound on $\mathcal{I}[m_u^H]$ in Eq.~(\ref{eq:bound2}), one obtains the bound on the up quark Yukawa at the scale of $m_Z$:
\begin{equation}
|Y_u  (m_Z) | <  0.05 \ Y_\nu  \left(\frac{0.1}{\sin \theta^H_{\rm QCD} } \right),
\label{eq:Yu}
\end{equation}
where we have taken into account the running of the up quark mass due to QCD interactions $|m_u(m_Z)|/|m_u (1 \GeV)| \sim 0.4$. For the $SU(3)^3$ instanton configuration~\cite{Agrawal:2017evu}, the required vanishing tree-level Yukawa coupling of strange and bottom quarks to the
$H$ and $\Phi$ Higgs fields may be simply ensured by assigning $s_R$ and $b_R$ the same $\mathcal{Z}_4$ charge as the one for~$u_R$.
 
As pointed out in Ref.~\cite{Agrawal:2017evu}, after the generation of the proper CKM mixing angles, 
one obtains flavor violating effects that demand the $SU(3)^3$ breaking scale to be larger than a few 100's~of~TeV.  Moreover, the corresponding off-diagonal Yukawa
couplings lead to instanton corrections to the imaginary component of the quark masses. These corrections modify the value of $\theta_{\rm QCD}$ at the $SU(3)^3$ instanton scale,
and, if they are evaluated at the scale $\Lambda_i \sim { \cal O}({\rm few~}100~{\rm TeV})$, they are of the order
of $10^{-11}$, and hence an order of magnitude smaller than the current bound on $\theta_{\rm QCD}$.
One potential problem of the formulation presented is that the hierarchy between $m_\Phi$ and the electroweak scale is not stable in the presence of $\lambda_{\Phi,1},\lambda_{\Phi,2},\rho$.  
To address this problem, in the next section we present a supersymmetric extension of  this scenario.  \\

%====================================================================%

\noindent\underline{\textbf{\textit{Supersymmetric Extension.}}} 
In the case of Supersymmetry~(SUSY), we assume the presence of a $\mathcal{Z}_3$ symmetry, and charges $\Phi_u: -1, \quad \Phi_d: 1, \quad u^c_R: 1, \quad \nu^c_R: 1$, $S:-1.$ All other fields are neutral under  the discrete $\mathcal{Z}_3$ symmetry.  The corresponding superpotential is given by
\bea
W & = & -Y_\nu^* L \Phi_u \nu^c_R -  Y_u^* Q \Phi_u u^c_R 
-  y_e^* L H_d e^c_R 
- y_d^* Q H_d d^c_R  \nn \\
&+&  \mu H_u H_d +m_\Phi \Phi_u \Phi_d + \lambda H_u \Phi_d S + \frac{\kappa}{3} S^3 .
\eea
 Right-handed neutrino Majorana masses generated by the singlet $S$ are forbidden by the holomorphicity of the superpotential. In addition, we have imposed  $R$-parity which forbids terms like $(v^c_R)^3$.   The SUSY invariant potential for the Higgs fields reads:
\bea
V_{\text{SUSY}} &=& |\mu|^2 |H_u|^2 + |\mu H_d + \lambda\Phi_d S|^2+   |m_\Phi \Phi_u + \lambda H_u S|^2\nn \\
 &+& |m_\Phi|^2| \Phi_d  |^2 +  | \kappa \ S^2 + \lambda H_u \Phi_d| ^2 ,
\eea
where $\mu$ is  the conventional $\mu$ term. In the following, we will take $m_\Phi \gg \mu \sim $ TeV. After  SUSY-breaking, we have the following soft-breaking interaction terms:
\beq
\begin{split}
V_{\rm soft} &=m_{\Phi_u}^2 |\Phi_u|^2 +  m_{\Phi_d}^2 |\Phi_d|^2 + m_S^2 S^* S +\cdots\nonumber\\
&+ ( \lambda a_\lambda  H_u \Phi_d S+ b_\lambda \Phi_u^\dagger H_u S  + a_\kappa S^3 + \cdots + h.c.),
\end{split}
\eeq
where we have omitted  terms not relevant for our discussion. Note that in the limit of $\kappa = a_\kappa = 0$ there would be a  $U(1)$ global symmetry which would make the singlet CP-odd scalar massless. More specifically, a global $U(1)$  Peccei-Quinn symmetry~\cite{Peccei:1977hh} is broken by the $S^{2} (H_u \Phi_d)^*$ and $S^3$ terms, which are proportional to $\kappa \lambda$ or $a_\kappa$. Since $\Phi_d$ acquires a very small vev, the mass of the CP-odd scalar  predominantly originates from a negative $a_\kappa$.

By assuming that the  SUSY-invariant  mass $m_\Phi$ is much larger than all the soft masses, one can integrate out the  heavy scalar fields $\Phi_{u,d}$ :
$\Phi_u \sim - \frac{\lambda}{m_\Phi} H_u S$,  $\Phi_d \sim -\frac{1}{|m_\Phi|^2}\left(\mu \lambda^* H_d S^* +  \lambda a_\lambda  \tilde H_u S^*  \right)$, and obtain the low-energy effective Lagrangian for the Yukawa interactions  in the Dirac fermion notation:
\bea
\mL^y_{\rm eff} & =  & -Y_\nu \frac{\lambda^* S^*}{m_\Phi^*} \bar{\ell}_L \tilde{H}_u  \nu_R   - Y_u \frac{\lambda^* S^*}{m_\Phi^*} \bar{q}_L \tilde{H}_u u_R + \cdots
\eea
from which we can read off the neutrino and up quark masses:
\beq
\label{eq:mnumu}
m_\nu \sim \left(Y_\nu \frac{\lambda^* v_S^*}{\sqrt{2} m_\Phi^*} \right) \frac{v_u}{\sqrt{2}}, \quad m_u^H \sim \left(Y_u \frac{\lambda^*  v_S^*}{\sqrt{2} m_\Phi^*} \right) \frac{v_u}{\sqrt{2}},
\eeq
where we assumed $v_u$ to be real, and the expression between parenthesis on the left- and right-hand side of Eq.~(\ref{eq:mnumu}) defines the low energy Yukawa couplings $y_\nu$ and $y_u$, respectively. The necessary values  of $|m_\Phi|$ and 
$|Y_{u,\nu}|$ can be extracted from Eqs.~(\ref{eq:mphi})~and~(\ref{eq:Yu}) after replacing $|\rho/m_\Phi|$ by $|\lambda|$, and $v$ by $v_u$.   For the $SU(3)^3$ case, as in the non-SUSY scenario, the required vanishing tree-level Yukawa coupling of strange and bottom quarks to the $H_d$ and $\Phi_{d}$ Higgs fields may be simply ensured by assigning $s_R^c$ and $b_R^c$ the same $\mathcal{Z}_3$ charge as the one for $u_R^c$.

Generically supersymmetric extensions lead to additional contributions to the electric dipole moments.  In the absence of flavor violation in the scalar mass parameters, they are proportional to the phases   $\Phi_A^{if} = \arg [M_{i}\, A^*_f], \Phi_B = \arg[M_{\tilde{g}}^* \,\mu^* \, (B\mu)]$, where
 $y_f A_f$ are the scalar trilinear couplings, $M_{\tilde g}$ is the mass of the gluino, $M_i$ the gaugino masses, and  $B\mu$ the $H_uH_d$ bilinear mass parameter. The one-loop SUSY corrections to the nEDM, controlled by  $\Phi_A^{if}$ and $\Phi_B$, may be parametrized as~\cite{Kolda:1997wt}
 \begin{equation}
 \label{eq:dnSUSY}
 d_n^{\rm SUSY}  \simeq 2 \left(\frac{100 \ {\rm GeV}}{m_{\rm SUSY}}\right)^2 \Phi_{A,B}^{if}10^{-23} {\rm e \ cm},
 \end{equation}
 where $m_{\rm SUSY}$ denotes a common soft supersymmetry breaking mass scale.
There are also relevant contributions at the two-loop level, that lead to a somewhat more complicated dependence on the SUSY and Higgs spectrum, as well as to possible cancellations between one and two loop contributions~\cite{Ibrahim:2007fb,Ellis:2008zy}. These contributions will be suppressed well below the current bounds without fine-tuning the CP-violating phases if the masses of the gluino, squark and heavy Higgs boson masses are  larger than 10~TeV.

 An important consideration is that  after integrating out the SUSY particles, 
 the low energy Yukawa couplings are affected by non-decoupling and CP violating contributions, proportional to $\Phi_{A,B}^{if}$~\cite{Carena:2000yi}.  
 Hence, if the instanton scale is above the supersymmetric particle mass scale, the proposed solution to the strong CP problem will be invalidated by the appearance 
 of new phases in the Yukawa couplings. Therefore, we must demand the supersymmetry particle masses to be above the instanton scale. Moreover, for the up-quark 
 Yukawa coupling to remain small after supersymmetry particle corrections, we should
 demand that the supersymmetry breaking mechanism preserves the $\mathcal{Z}_3$ symmetry.   
  
 If the instanton effects come from regular $SU(3)$ interactions,  the supersymmetry particle and heavy Higgs boson masses are naturally much larger than the QCD instanton scale.     However,  for the $SU(3)^3$ scenario our proposed solution of the strong CP problem is only viable  if heavy Higgs and colored SUSY particle  masses  are of the order of or larger than the characteristic $SU(3)_i$ instanton scales $\Lambda_i$.   As discussed above, $\Lambda_i$ must be $\sim \mathcal{O}$(few 100 TeV)~\cite{Agrawal:2017evu}. This suppresses all the CP-violating and flavour-changing effects induced by the heavy Higgs and SUSY particles in Eq.~(\ref{eq:dnSUSY}). On the other hand, it introduces a little hierarchy problem, which will not be  addressed further in this work.\\

%====================================================================%

\noindent\underline{\textbf{\textit{Neutron Electric Dipole Moment.}}} 
A notable outcome of our framework is that a non-zero nEDM  is induced by the non-vanishing value of $\theta_{\rm QCD}$. 
%====================================================================%
 \begin{figure}
	\centering
    \includegraphics[width=0.7 \columnwidth]{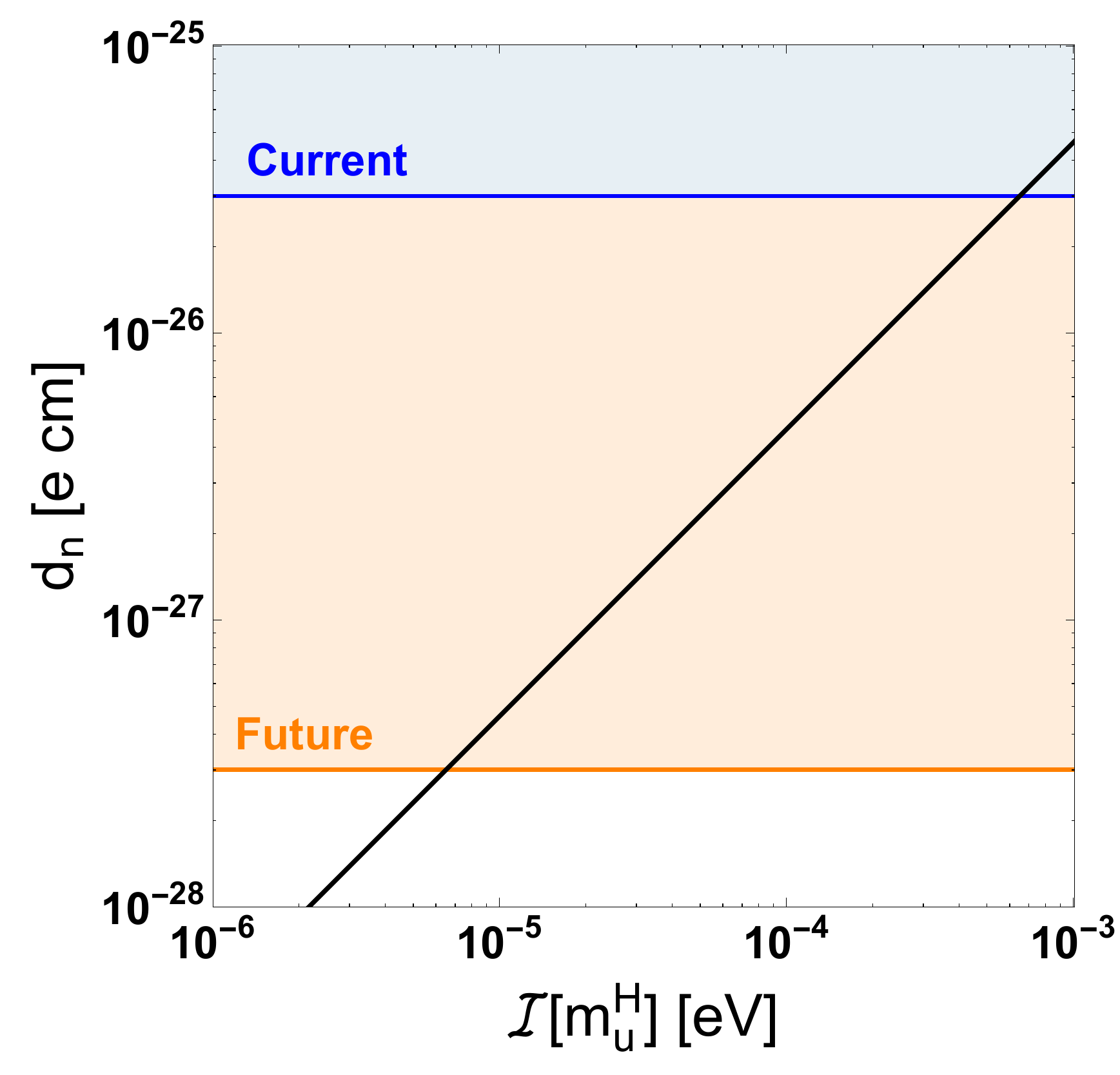} 
	\caption{The neutron EDM as a function of the imaginary part of the up quark mass. We have also shown the current 90\% C.L. bound~\cite{Afach:2015sja}  and prospective sensitivity from the future neutron EDM measurements~\cite{Abel:2018yeo,Picker:2016ygp,Schreyer:2018lih,Slutsky:2017mbn, Serebrov:2017sqv,Ito:2017ywc,Kuchler:2016eik}.}
	\label{fig:neutronEDM}
\end{figure}
%====================================================================%
We can calculate the contribution to the nEDM from current algebra~\cite{Crewther:1979pi,Baluni:1978rf}; the result reads:
\beq
\frac{d_n}{e} \sim  \frac{g_{\pi N N} ~ \bar{g}_{\pi N N}}{4 \pi^2 M_N}  \ln \frac{M_N}{m_\pi},
\eeq
where $M_N \sim 940$ MeV is the nucleon mass, $m_\pi \sim 140$~MeV is the pion mass and  $|g_{\pi N N}| \sim 13.4$ is the usual CP conserving pion-nucleon coupling. The CP  violating coupling $\bar{g}_{\pi N N}$ is given by:
\bea
\bar{g}_{\pi N N} &\sim& \theta_{\rm QCD}  \frac{m_{\rm eff}}{F_\pi },
\eea
with  
$m_{\rm eff}  \equiv  |m_um_dm_s|/(|m_um_d| + |m_um_s| + |m_dm_s|)$, 
$F_{\pi} \sim 93 \,\text{MeV}$ is the pion decay constant, and the masses of the quarks and the strong CP phase are evaluated at the scale $Q \sim 1$ GeV. Using the currently determined values for $|m_{u,d,s}|$~\cite{Tanabashi:2018oca}, this result becomes consistent with the calculation of Refs.~\cite{Pospelov:1999mv,Pospelov:1999ha} by using  the QCD sum rules,
\beq
d_n \sim \theta_{\rm QCD} \times (2.4 \pm 0.7) \times 10^{-16} \,{\rm e \ \text{cm}},
\eeq
and also with a recent lattice calculation~\cite{Dragos:2019oxn}. In the canonical basis, where $  \theta_{\rm QCD}  \sim  \mathcal{I}[m_u^H]/|m_u|$, and normalizing the value of the nEDM to the  present  bound~\cite{Afach:2015sja},  we obtain
\beq
\label{eq:dnbound}
d_n  = \frac{\mathcal{I}[m_u^H]}{(6.5  \pm 2.0) \times 10^{-4} {\rm eV}} \times 3.0 \times 10^{-26} {\rm e ~ cm}.
 \eeq

Figure~\ref{fig:neutronEDM} shows the nEDM as a function of the imaginary part of the up quark mass. While the current measurement leads to a bound on $\mathcal{I}[m_u^H] < (6.5 \pm 2.0)\times 10^{-4}$~eV, future nEDM experiments~\cite{Abel:2018yeo,Picker:2016ygp,Schreyer:2018lih,Slutsky:2017mbn, Serebrov:2017sqv,Ito:2017ywc,Kuchler:2016eik} will be able to improve the present sensitivity by two orders of magnitude $\sim 3 \times 10^{-28} \, \text{e  cm}$~\cite{Slutsky:2017mbn}, and hence will be able to probe $\mathcal{I}[m_u^H]$ up to  about  $6 \times 10^{-6}$~eV.  Note that  even for a phase $\theta^H_{\rm QCD} \simeq 10^{-2}$, the values of $|m_u^H|$ that will be probed are much smaller than the ones that naturally arise from the relation of $m_u^H$ and the neutrino masses. Hence, it is natural to expect a measurement of the nEDM by the next generation of experiments within this framework. 

Finally, we should comment on additional  contributions to the nEDM. As discussed above, they can either come from sources of CP violation associated with the new physics introduced to
stabilize the scale hierarchies, Eq.~(\ref{eq:dnSUSY}),
or, in the $SU(3)^3$ scenario~\cite{Agrawal:2017evu}, from instanton  contributions to the imaginary part of the quark masses, arising after the generation of off-diagonal Yukawa couplings. 
While the former are suppressed by the square of the new particle masses, the latter are about an order of magnitude smaller than the current bound on the nEDM.
Although these corrections may potentially break the correlation
between the nEDM and the neutrino masses, barring an unlikely strong cancellation, they reinforce the expectation of a measurement of the nEDM in the near future.  \\
%====================================================================%
~\\
{\bf Acknowledgments:} We would like to thank P.~Agrawal, W. Jay, A. Kronfeld and L.T. Wang for useful discussions and comments. This manuscript has been authored by Fermi Research Alliance, LLC under Contract No. DE-AC02-07CH11359 with the U.S. Department of Energy, Office of Science, Office of High Energy Physics. Work at University of Chicago is supported in part by U.S. Department of Energy grant number DE-FG02-13ER41958. 
Work at ANL is supported in part by the U.S. Department of Energy under Contract No. DE-AC02-06CH11357. NRS is supported by Wayne State University and by the U.S. Department of Energy under Contract No. DESC0007983. JL acknowledges support by an Oehme Fellowship. 
%====================================================================%

\bibliographystyle{apsrev}
\bibliography{references}

\end{document}